\documentclass{article}

\usepackage{PRIMEarxiv}

\usepackage[utf8]{inputenc} % allow utf--8 input
\usepackage[T1]{fontenc}    % use 8--bit T1 fonts
\usepackage[authoryear]{natbib}  % enable author--year citations
\usepackage{hyperref}       % hyperlinks
\usepackage{url}            % simple URL typesetting
\usepackage{booktabs}       % professional--quality tables
\usepackage{amsfonts}       % blackboard math symbols
\usepackage{nicefrac}       % compact symbols for 1/2, etc.
\usepackage{microtype}      % microtypography
\usepackage{fancyhdr}       % header
\usepackage{graphicx}       % graphics
\usepackage{amsmath,amssymb,amsthm}
\usepackage{subcaption,pgfplots,float}
\usepackage{xcolor}
\usepackage{lipsum}
\usepackage{bm}
\newcommand{\permil}{\text{\textperthousand}}

\graphicspath{{}}

\title{  Bayesian Synchronization of Proxy Paleorecords with Reference Chronologies
}
\author{
Marco A. Aquino--L\'opez$^{\S 1}$~Francesco Muschitiello$^{2,3}$~Matthew Osman$^2$ \\
$^\S$\texttt{Corresponding Author: aquino@cimat.mx} \\
\normalsize{$^1$ Centro de Investigaci\'on en Matem\'aticas (CIMAT), Guanajuato, M\'exico} \\
\normalsize{$^2$ Department of Geography, University of Cambridge, United Kingdom} \\
\normalsize{$^3$ Centre for Climate Repair, University of Cambridge, Cambridge, UK}
}

\begin{document}

\maketitle

\begin{abstract}
Many scientific fields compare two or more noisy time series that integrate the same underlying process but are recorded on different time scales. In paleoclimate studies, for example, proxy measurements are collected versus stratigraphic depth in a climate archive and then converted to calendar time. Synchronizing two proxy records often requires estimating an \emph{alignment} that maps the depth (or preliminary age) of an \emph{input} record onto the calendar--time scale of an absolutely--dated \emph{target} record so that corresponding proxy signals line up. Existing alignment approaches are generally optimization--based and return a single transformation, providing limited formal uncertainty quantification.

Here, we introduce \texttt{BSync}, a Bayesian synchronization framework that treats alignments as inference over a monotone time--mapping function to match an input to a target record. The alignment is expressed as a transformation of the input depth (or age) scale to match the target record, achieved through a link function that locally expands and compresses the input scale. The model is parameterized through interpretable local rate parameters, enabling the specification of priors on deposition times to regularize the alignment toward physically plausible deformations. \texttt{BSync} jointly infers the aligned chronology and provides posterior uncertainty for the time--warping function and the resulting age scale. In synthetic data experiments and a real--data case study, \texttt{BSync} yields well--calibrated credible intervals for the aligned time scale and achieves more accurate alignments than a state--of--the--art automated method, particularly when independent age constraints are sparse. 
\end{abstract}
	\noindent%
	\vfill
	\newpage
	% \spacingset{1.45} % DON'T change the spacing!

\section{Introduction}

Sedimentary sequences recovered from the ocean floor, lakes, peat bogs, and other depositional environments preserve physical, biological and chemical fingerprints whose variations are often linked to past climate and environmental changes. Since direct observations of key climate variables are only available for recent decades --- a period covering less than $10^{-7}$ of Earth's history \citep{bradley2015} --- these fingerprints, or ``proxies", are our only window into Earth's deep climate history.  The stratigraphic alignment of these climate proxy profiles is thus an essential application in for paleoclimate research: by matching the distinctive ``wiggles" found across different proxy records, researchers can place multiple climate archives onto a shared, internally consistent age scale.  This permits environmental signals to be robustly compared across different regions and environmental domains well beyond the instrumental era.  

\begin{figure}[htbp]
\centering
\includegraphics[width=0.7\textwidth]{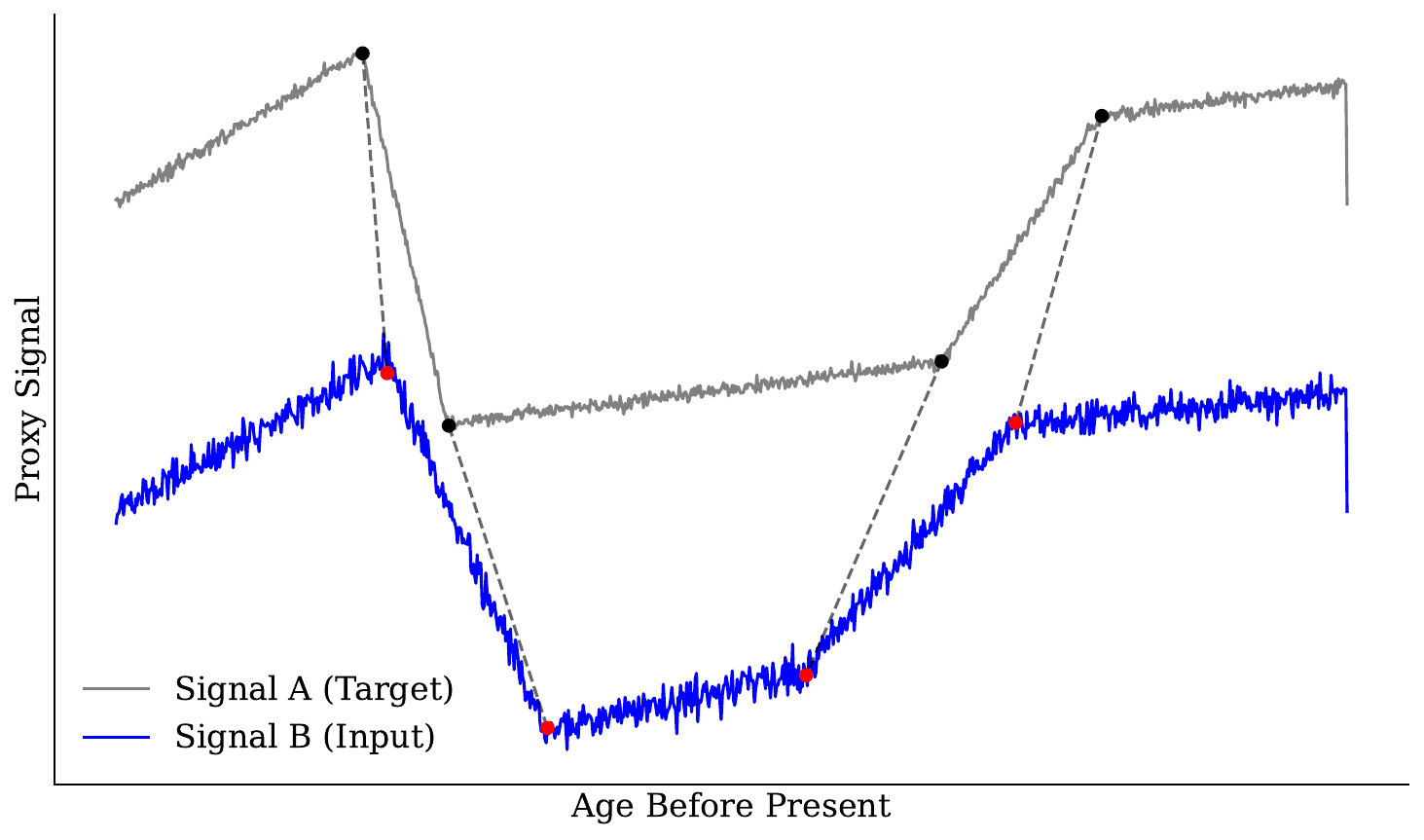}
\caption{Illustration of stratigraphic alignment based on ``manual'' synchronization of common transitions (i.e., ``tie--points'') determined between two climate proxy signals. The gray curve (Signal A) serves as an absolutely dated target record, while the blue curve (Signal B) represents the input record placed on a preliminary age scale before synchronization. Each record is divided into a set of stratigraphic periods, shown as horizontal bars: black bars for Signal A and red bars for Signal B. These bars represent the time intervals between successive transitions, and the midpoint of each interval defines the tie--point used for alignment. Dashed lines connect corresponding midpoints where the timing of both signals is assumed to be synchronous.
    	}\label{fig:sync_ilustration}
\end{figure}

The stratigraphic alignment (or synchronization) of climate records has historically been pursued by manually identifying and matching tie points marking sharp or distinctive transitions in the proxy stratigraphies in both an undated input record and an absolutely--dated target record (e.g.\citep{Shackleton_1973}). Figure~\ref{fig:sync_ilustration} shows an illustration of this practice, which assumes that climate transitions are synchronous in both input and target proxy signals.
This tie--point synchronization is especially prevalent in paleoceanographic studies, where the lack of independent chronological constraints can hinder the construction of precise timescales \citep{Muschitiello2020}, particularly beyond the radiocarbon calibration window (ca.\ 50~ka before  1950~AD; \citep{Reimer2020IntCal20,Blaauw2011}). Even when distinct proxy features exist --- for example, abrupt climate shifts such as Dansgaard--Oeschger events \citep{Hughen_2006,Capron2021} --- this approach remains limited because it relies on few discrete stratigraphic markers, which inhibits continuous quantification of alignment uncertainties. Additionally, tie--point alignments introduce subjective constraints on sedimentation, often resulting in ``step changes" in deposition times that are difficult to validate or reconcile with realistic sedimentation processes. The alignment process is further complicated by the fact that sediments compact, expand, and accumulate unevenly over space and time \citep{Sadler_1981}, resulting in markedly different accumulation histories across sites and depositional environments, regardless of proxy data resolution, or degree of proxy ``noise".

% -------- Figure 1 -----------

Despite the circularity that climate synchronization entails (e.g. \citet{Blaauw_2012}), proxy alignment remains a widely used and often necessary step in paleoclimate research. In fact, there is an urgent need to develop alignment methods that minimize subjectivity, ensure reproducibility, and allow quantifying uncertainties associated with the alignment process. In this study, while acknowledging the potential for interpretive circularity, we assume that tuning itself does not compromise the integrity of results; we thus focus on developing a more robust and reproducible alignment tool for geophysical problems where stratigraphic alignment is required.

%Deterministic methods 
Beyond tie--point synchronization, more objective methods have been developed to estimate alignments by means of nonlinear optimization \citep{Brueggemann1992,Loizeau1997,Lisieki2002,Marwan2002,LisieckiRaymo2005,LisiekiHerbert2007}. These approaches typically maximize an overall goodness--of--fit metric by deforming the entirety of a proxy time series onto a target record. However, because the optimization is subject to constraints and penalties designed to impose sedimentation realism and guard against undesirable fitting behavior, these methods do not provide a strictly formal quantification of alignment uncertainty.  Attempts have been made to quantify alignment uncertainty in a continuous fashion (e.g., \citet{Malinverno2013}). However, these methods often entail subjective and time--consuming ``link functions", where the depth scale of the input stratigraphy must be aligned to a target record via manual selection of tie points prior to alignment. 

Over the past decade, significant progress has been made in addressing these limitations using Bayesian techniques. The benefits of these methods are manifold: they produce physically grounded alignments based on prior expectations of sediment accumulation rates variability (\citep{Lin2014,Lee2023,Muschitiello_2024}), and provide credible intervals associated with the alignment process.

However, despite their clear potential, these methods still have limitations. For instance, most have been designed for aligning benthic foraminifera--derived $\delta^{18}$O data, which typically have low temporal resolution and high signal--to--noise ratios. The most recent alignment method --BIGMACS-- \citet{Lee2023}, uses Gaussian Process regression; while powerful, this technique incorporates smoothing techniques that may dampen important information in the proxy data. Consequently, uncertainty estimates may be underestimated in certain cases, leading to overconfidence in the precision of the alignment. In addition, \texttt{BIGMACS} uses an empirical sedimentation prior based on a limited set of radiocarbon dates from low--latitude ocean sediment cores. Although this approach may work well in some contexts, the empirical prior reflects assumptions specific to those environments that do not account for radiocarbon reservoir effects and are entirely absent for polar oceans, so it may misrepresent the alignment characteristics of the sediment core being analyzed.

To address some of these limitations and to improve the uncertainty quantification of the alignment process, here we present a new Bayesian synchronization method named \texttt{BSync}. \texttt{BSync} provides an integrated and generalized approach for aligning proxy time series from various sedimentary archives, including deep--sea sediments, lake sediments, peat bogs, and aeolian deposits.  \texttt{BSync} uses a Bayesian approach to estimate alignment uncertainty, explicitly modeling the non--climatic ``noise" inherent to proxy data. Furthermore, \texttt{BSync} incorporates a flexible sedimentation model that permits bespoke prior knowledge about deposition times (which we parameterize using \texttt{acc\_mean}, following the notation of \citep{Blaauw2011}) as ``priors" in the model, improving the model's ability to perform across different depositional environments. In addition, \texttt{BSync} offers the novel capability of aligning an input signal to a mixture of two chronologically consistent target time series. This methodology is implemented as an open--source synchronization routine written in the programming language \textit{R}, offering the possibility for future customization and extensions.

The remainder of this paper is organized as follows. Section~\ref{sec:model} introduces the \texttt{BSync} framework, its three alignment strategies, likelihood choices, priors, and the Markov Chain Monte Carlo (MCMC)-based implementation. Section~\ref{sec:classical_ali} puts the method to work using marine sediment cores, showing how \texttt{BSync} handles benthic and planktic $\delta^{18}$O records under different alignment scenarios. Section~\ref{sec:comp_exp} turns to controlled computational experiments, evaluating bias, coverage, and interval width across a range of noise levels, sampling resolutions, and mixed--target cases. Section~\ref{sec:bigmacs} provides a comparison with the current state--of--the--art alignment algorithm, \texttt{BIGMACS}, outlining the differences in modeling assumptions, priors, and data--noise characteristics between \texttt{BIGMACS} and \texttt{BSync}. Finally, Section~\ref{sec:discussion} synthesizes the findings, discusses limitations, and points to possible future extensions for Bayesian alignment in paleoclimate research.

%%%%%%%%%%%%%%%%%%%%%%

\section{Models}\label{sec:model}

The core objective of \texttt{BSync} is to estimate the deformation of an input record that maximizes the alignment to one or more target records, while formally accounting for prior knowledge on deposition times associated with this process. 

\texttt{BSync} offers three distinct alignment strategies between an input and a target proxy signal, suitable for cases where the input chronology is reliable \citep[see:][]{Muschitiello_2024} and where the input record lacks an age scale. These alignment options are: single target alignment (classical alignment problem), double target alignments (alignment to a mixture of two targets) and alignment to a target with reported age uncertainty obtained from e.g. a Bayesian age--depth model. Although each strategy has specific considerations for constructing the likelihood and parameter space, the data pre--treatment and link function remain consistent across strategies. Each strategy can be used to improve an existing age--scale or to create a new one, with minor modifications to the parameter space and prior distributions. 

The core component of \texttt{BSync} is a link function $\tau$ that maps the input record onto the target age scale, in a way that maximizes the fit between the input and target records. Formally, this function takes the form $\tau: D \rightarrow T'$ or $\tau: T \rightarrow T'$, where the domain $D \subset \mathbb{R}^+$ represents depth values (when the input record is in depth), and $T \subset \mathbb{R}$ represents calendar ages (when the input record has already been placed on an age scale). The shared range set $T'$ corresponds to the age scale of the target record(s). Assuming that both input and target ages increase monotonically as depth from the surface increases (i.e. sediments accumulate material over time, leading to older material being found in deeper parts of the sedimentary record), the function ($\tau$) should be bijective.

Because \texttt{BSync} relies on matching similar structures between input and target proxy records, it is essential that the time range of the input record ($T$) is fully contained within the time range of the target ($T'$), i.e., $T \subseteq T'$. This ensures that any transformation of the input scale used to align features will be within the bounds of the target, preventing extrapolation beyond the target data. We assume that users have some knowledge of the age range of the input record either via preliminary analysis and (or) visual identification of key features in the record.

\subsection{Data pretreatment}\label{sec:rescale}

To reduce the number of parameters the model needs to infer, \texttt{BSync} implements a pre--processing phase in which both input and target proxy values are scaled. This approach eliminates the need for additional parameters that would otherwise be required to bring the records to comparable scales.

We use a non--parametric quantile rescaling to map specified lower and upper quantiles ($q_l,q_u$) within the $[-1,1]$ interval, ensuring that the inter--quantile range of the data is standardized across datasets, this approach was explored in detail in \cite{Muschitiello_2024}. The rescaled value $x'$ of a given proxy observation $x$ is computed as:
\begin{equation}\label{eq:rescaling}
x' = \frac{2(x - x_l)}{x_u - x_l} - 1
\end{equation}
where $x_l$ and $x_u$ denote the values corresponding to the lower and upper quantiles in the original record. This quantile rescaling approach ensures that both records are comparable, minimizing the impact of outliers. While this method is not without limitations and may still be sensitive to extreme values, we encourage users to evaluate their proxy data and adopt a suitable transformation/scaling strategy tailored to their specific needs prior to alignment.

\subsection{Target considerations}
We now turn our attention to the characteristics of the target record. Since the target is assumed to have a reliable age scale, it plays a central role in the alignment process. However, the age of the target records may be reported in different forms. In some cases, the target is provided with a ``trusted" age scale, often the ensemble mean or median of an age model, with no associated uncertainty. In more ideal scenarios, it is accompanied by posterior samples from a Bayesian age--depth model. The user may also be interested in aligning the input to a combination of target records in cases where multiple candidate target records are available. To accommodate these possibilities and make full use of the available age information, we overview three alignment strategies tailored to each scenario.

\subsubsection{Single target}

The ``single target" approach corresponds to a classical alignment problem, where the age of the target record is treated as fixed and known, with no associated uncertainty. In this scenario, \texttt{BSync} performs the alignment in a manner similar to that proposed by \citet{Muschitiello_2024}, treating the scaled proxy values of the target as the ``true" values to which the input record is aligned. In short, the stratigraphic scale of the input record (either on depth or age) is systematically deformed until the proxy signals align with the structure of the target proxy signal. The likelihood function used for this approach is given in Equation~\ref{eq:like_single}, Section~\ref{sec:Likelihood}.

\subsubsection{Mixing of two targets}

\texttt{BSync} includes an optional feature that allows alignments against two distinct target records, $v_1$ and $v_2$, through a mixing parameter $\omega_m$. This novel ``mixing model" approach can be understood through both spatial and dynamical contexts. Spatially, an input site might exist along a climatological gradient between two previously aligned target sites; in this case, the mixing model approach combines the two target records to optimally reflect the input site's intermediate position. Dynamically, the mixing model can represent the interplay between opposing ``push" vs. ``pull" climatic forcings \cite{Rohling2009}, such that conditions at the input proxy site are conditioned by both.  In either framework, the mixing model ``tunes" the input record against multiple established target records by optimizing the contribution of each site's influence. This approach mitigates against site--specific noise by combining two well--constrained and chronologically coherent targets, and is particularly advantageous when both datasets offer complementary information and strength or when no singularly optimal alignment target is apparent.

The two signals are combined according to the following formula:
\begin{equation}
	v = \omega_m v_1 + (1-\omega_m) v_2.
\end{equation}

This approach assumes that both target records ($v_1$ and $v_2$) are in consistent units, ensuring that their structural similarities are meaningful for comparison.  In this process, we first create a mixed target signal, denoted by $v$, which combines the original signals from two distinct target records, $v_1$ and $v_2$, each assumed to have reliable age scales. Following this mixing step, we then rescale $v$, as described in section \ref{sec:rescale}, to function as the adjusted target signal in the alignment. The parameter $\omega_m$ is a mixing parameter that controls the weighting of $ v_1 $ and $ v_2 $ in the mixing process; it has a domain of $(0,1)$. This formulation allows the model to derive a posterior distribution for $\omega_m$, thus providing a posterior sample of the target signals that reflects the combined characteristics of both records.

Physically, the mixing parameter $\omega_m$ represents the relative climatological influence of each target record on the mixed target signal.  $\omega_m$ allows the model to adjust the contribution of each signal to the paleoclimatic conditions at the input site. By applying this mixing process, the model not only reduces the effects of multiple signals and noise but also provides insight into which of the two target signals is most similar to the signal being aligned. This feature can potentially be used to study the control of different climatic signals in different regions of the planet \citep{Rohling_2009,Barker_2007}, assuming stationarity in the relative influence of $v_1$ and $v_2$ on $v$.

Once the new target is created, similar assumptions as those used in the single target approach can be made, by considering the mixed target's scaled values as the ``true'' values for the input record. The likelihood associated with this approach is presented in equation \ref{eq:like_double}, Section \ref{sec:Likelihood}.

\subsubsection{Reported age uncertainty in the target}

If a target record's age scale is derived from a Bayesian age--depth model or comparable approach, neglecting the inherent uncertainty would not only be inaccurate but would also disregard valuable information offered by the posterior distribution. Recent methodologies (e.g. \citep{McKay_2021}), aim to integrate age uncertainty directly into paleoenvironmental analyses. This is achieved by leveraging all iterations from the Markov Chain Monte Carlo output of Bayesian age--depth models, performing analyses across each iteration to account for this uncertainty. However, for our specific case, this approach is infeasible due to its excessive computational demands, which would require conducting numerous single--target alignments.

Instead, we favor the approach taken by \citet{Aquino_2024}, which generates a non--parametric kernel estimator of the posterior distribution of the proxy at a given age. This allows us not to replace point estimates of the target record values with an estimation of the full posterior distribution. By using the posterior samples from the age--depth model,  \texttt{BSync} can incorporate age uncertainty into the alignment process, providing a more comprehensive handling of chronological uncertainties compared to traditional approaches that rely on singular age estimates. The likelihood corresponding to this methodology is detailed in equation \ref{eq:like_uq}, which can be found in Section \ref{sec:Likelihood}.

\subsection{link function}\label{link}

To maximize the similarity between the input and target records, it should be noted that the use of this methodology implicitly assumes that the two time series share some common underlying structure; otherwise the procedure may simply align noise with noise. Therefore, we must define a link function between the two records. The target record is assumed to provide a reliable age scale, whereas the input record either lacks an age scale or has a preliminary age--depth model that we aim to improve through alignment. This leads to two possible alignment strategies: (1) The input has no age scale, i.e. the link function ($\tau$) should be an age--depth relationship; (2) the input's age--scale requires adjustment, i.e. the link function is an age--to--age relationship.

In the first scenario, we use the inverse--gamma model proposed by \citet{Blaauw2011} to build $\tau$. This age--depth relationship is well--established in the paleoecological and paleoceanographic community (e.g. \citet{muglia2023,goring2012}), and in our current approach it provides a series of benefits that naturally aid the alignment process: for example, it permits variable sedimentation rates, and incorporates uncertainty directly into the age--depth relationship.  The inverse--gamma model requires the input depth scale $D$ to be divided into sections of equal length ($\Delta c$), thereby dividing $D$ into $K$ subsets ($[c_{i--1},c_i)$). Let $\tau_0$ be the age of the uppermost core section, and let $\textbf{m}$ be a vector of length $K$ describing the slopes of the $K$ sections in the input depth domain. With this notation, we can define the age of $x$ as:
\begin{equation}
\tau(d) = \tau_0 + \sum_{i=1}^{j} \left( m_i\Delta c \right) + m_{i+1}(d - c_i),\label{eq:tao}
\end{equation}
where $d$ is a given depth such that $d\in D$. Note that because $\tau(d) \in T'$ $\forall$ $d\in D$, we have $\tau_0\in T'$. Moreover, this restriction provides a well--defined search domain. In this particular instance, $m_i$ is interpreted as the deposition time in the input environment. This interpretation gives us a natural restriction as $m_i > 0$ $\forall$ $m_i \in m$.

For the second case, where the initial age scale $T$ requires refinement and the link function serves to adjust it, we can substitute the depth space for the age space $T$, and thus equation \ref{eq:tao} can be rewritten as:
\begin{equation}
\tau(t) = \tau_0 + \sum_{i=1}^{j} \left( m_i\Delta c \right) + m_{i+1}(t - c_i),\label{eq:tao2}
\end{equation}
where $t\in T$ is the original age scale, $\tau(t)$ is the corrected age of $t$, and $\tau_0=\tau(t_0)$ is the initial shift of the original age scale, and so $\tau_0\in T'$. In this case, $m_i$ does no longer reflect deposition times, but rather expansion/contraction ratios relative to the initial age scale. For this particular implementation, the initial age--depth model can be trusted to some extent, and we limit the possible values of $m_i$ such that $\frac{1}{4} < m_i < 4$ $\forall$ $m_i \in m$ , which we deem as a conservative estimate of relative sedimentation rate change (e.g. \citet{Muschitiello_2024,Lin2014}). This restriction means that we cannot compact or stretch the original age scale more than 4 times its initial age. This constraint permits plausible adjustments whilst preventing against extreme distortions of the initial age model.

Parameters $\mathbf{m} = (m_1, m_2, \ldots, m_K)$ represent deposition times of each section $[c_{i--1},c_i)$, but they are not the actual model parameters. Instead, the parameters of the model are increments $(\alpha_i)$ that undergo a transformation to calculate the actual deposition times $m_j$, which is expressed through the following equation:
\begin{equation}
m_j = \omega m_{j+1} + (1-\omega) \alpha_j, \label{eq:accum_eq}
\end{equation}
where $\omega \in[0,1]$ is a memory parameter, and $\alpha_j$ are positive values. This mechanism allows for a dynamic adjustment of the deposition times based on a combination of the deposition times from older sections ($m_{j+1}$) and new incremental data ($\alpha_j$). This means that the parameters used by \texttt{BSync} in both cases is $\tau_0,\alpha,\omega$.

\subsection{Likelihood}\label{sec:Likelihood}

Once we have pretreated the data to be in similar windows (both in the rescaled proxy and, where applicable, age scale) we need to define the likelihood.

To define our likelihood, we assume that the rescaled target record is the mean of the rescaled input record at a given age $\tau(x_i)$ and that the input record has a constant standard deviation ($\sigma$) throughout the record. We assume a constant standard deviation because, in most cases, records are reported without age nor analytical proxy uncertainty. Considering that this variable is unknown, $\sigma$ is treated as a random variable estimated by the model, with its prior distribution discussed in Section \ref{sec:prior}. This approach applies to both the single--target and double--target alignment strategies, although it differs for cases where age uncertainty is explicitly reported.

To evaluate the alignment, a probability distribution that encapsulates the uncertainty inherent to the alignment provided by the function $\tau(\cdot)$ is needed . To mitigate the impact of  non--climatic ``noise" as well as local deposition processes, we adopt a t--distribution, rather than a normal distribution \citet{christen2009}. This heavy--tailed distribution provides greater flexibility insofar that it reduces the impact of local signal variations and depositional noise present in the target, thus providing a more robust overall alignment. Within this framework, we assume the following:
\begin{equation}
u_t \mid a, b, \sigma, v_t, \tau \sim \mathcal{T}_{a,b}\left( v_\tau, \sigma \right)\label{eq:like_single}
\end{equation}
where $u_t$ is the rescaled input record at age $t$, $v_t$ is the target record at age $t$, $a$ and $b$ are the shape parameters of the t--distribution, and $\sigma$ is the standard deviation. In the case of the double target strategy, the likelihood will also depend on the weight parameter $\omega_m$ which generates $v_\tau = \omega_m v_\tau^1 + (1-\omega_m) v_\tau^2$, therefore,
\begin{equation}
u_t \mid a, b, \sigma, v_t, \tau, \omega_m \sim \mathcal{T}_{a,b}\left( v_\tau, \sigma \right).\label{eq:like_double}
\end{equation}

Most proxy records report only point estimates for their age scales, but some also include dating markers or the full output of a Bayesian age--depth model, such as MCMC samples from the posterior distribution. These posterior samples offer important information about both age estimates and their uncertainties. Because Bayesian models are a robust and well established tool for chronology construction, we adopt the approach of \citet{Aquino_2024} to incorporate age uncertainty into \texttt{BSync}. Specifically, we generate Monte Carlo samples of the proxy values corresponding to each target age. As demonstrated by \citet{Aquino_2024}, these draws represent posterior samples of the proxy at a given age. We then apply a non--parametric kernel density estimator to approximate their distribution, which in turn serves as the likelihood function in our framework.
\begin{equation}
u_t \mid v_t, i \sim \mathcal{K}_i\left( v_t \right)\label{eq:like_uq}
\end{equation}
where $ \mathcal{K}_i(x)$ represents the Gaussian kernel of the posterior sample at the $i$-th age--grid window containing $t$. In this case, the uncertainty around the proxy is assumed to be sufficiently large for the process that the parameter $\sigma$ is not necessary for this inference. Hereafter, we refer to the strategy based on this likelihood as UQ.

\subsection{Prior distributions}\label{sec:prior}

To complete the Bayesian model specification, we assign prior distributions to the model parameters. In our case, we have three main parameters to consider: the vector deposition times --in the case of the construction of an age--depth model-- or expansion/compression ratios --when improving an existing age model--  $m$, the initial age shift $\tau_0$, and the standard deviation $\sigma$.

\subsubsection{Prior for $\mathbf{m}$}\label{sec:accu_m}

The vector $\mathbf{m} = (m_1, m_2, \dots, m_K)$ represents the deposition times (or expansion/compression ratios) in the $K$ sections of the input record. Since theseare inherently positive, we assign independent Gamma prior distributions to the ``increments" $\alpha_i$:
\begin{equation}
\alpha_i \sim \text{Gamma}(a_i, b_i), \quad i = 1, 2, \dots, K.
\end{equation}
The Gamma distribution is a flexible choice that can accommodate a wide range of positive values, depending on the shape ($a_i$) and rate ($b_i$) parameters. These hyperparameters can be specified based on prior knowledge about the expected range and distribution of deposition times (or expansion/compression ratios). To help translate expert knowledge, \texttt{BSync}'s prior for $\bm{\alpha}$ is parameterized by the mean of $\bm{\alpha}$ and the shape parameter $a_i$. Instead, the deposition times are derived through a transformation involving both $\omega$ and the $\bm{\alpha}$ increments, as given by \ref{eq:accum_eq}. In cases where the BSync is employed to refine an initial age scale and the user opts for an age--to--age alignment approach, the global mean of the prior Gamma distribution defaults to $1$, implying no change from the initial age scale of the input record. As noted earlier, a restriction on $\bm{\alpha}$ is imposed, confining it to the interval $\left[\frac{1}{4}, 4 \right]$ (i.e., expansion or compression up to fourfold of the original age scale) to limit extreme deformations of the initial chronology \citep{Muschitiello_2024, Lin2014}.

To establish an appropriate prior for the parameters $\alpha$, we consider whether independent age--depth information is available and the degree of its associated uncertainty. Often, an initial age scale can be derived from independent chronological constraints like radiocarbon dates or tephra layers. Rather than including these dating markers in the likelihood function, we use the available age--depth model to inform the prior, thus translating it into priors for each $\alpha_i$.  This approach aligns with our age--to--age framework, as it allows flexibility in representing uncertainty by means of adjustable parameters ($a_i$, $b_i$) or global parameters derived from independent age constraints. In doing so, the initial chronology provides an informative starting point that is iteratively refined as new information is integrated, consistent with the principles of Bayesian learning.

In scenarios where no initial age scale is available and \texttt{BSync} must construct an age--depth model, the global prior mean deposition time is calculated using the minimum and maximum of $T'$ and $D$, such that $\mathbb{E}(m) = \frac{t'_n - t'_0}{d_n - d_0}$. Users are encouraged to adjust this default to reflect their expert prior knowledge about the sediments and the depositional environment.

\subsubsection{Prior for $\omega$}

The parameter $\omega$ serves as a memory parameter for the deposition time (or expansion/compression ratios) at each section $K$ of the model. It controls the degree to which the model retains the influence of the deposition time (or expansion/compression ratios) downcore, i.e. from the previous section.
We assign a Beta prior distribution to $\omega$:
\begin{equation}
\omega \sim \text{Beta}(\alpha_\omega, \beta_\omega),
\end{equation}
where $\alpha_\omega$ and $\beta_\omega$ are the shape parameters of the Beta distribution. In our case we use a mean of $0.5$ and a proportion (strength) of $\alpha_\omega + \beta_\omega = 10$, which provides a symmetric distribution around $0.5$. The Beta distribution is chosen due to its flexibility in modeling proportions, as it is defined on the interval $[0, 1]$  \citep{blaauw2012}. This makes it suitable for $\omega$, which must be within this range. This choice of priors for $\omega$ reflects a relatively uninformative stance, allowing the model to determine the influence of the previous rates relative to new increments, thus enabling a data--driven approach.

\subsubsection{Prior for $\tau_0$}

The parameter $\tau_0$ represents the initial age shift (or age offset) in the input record. Since $\tau_0$ is a location parameter on the age scale, we assign a Truncated Normal prior distribution to ensure $\tau_0\in T'$:

\begin{equation}
    \tau_0 \sim \text{TruncNormal}(\mu_{\tau_0}, \sigma^2_{\tau_0}, t'_0, t'_m),
\end{equation}
where $t'_0$ and $t'_m$ are the minimum and maximum ages of the target respectively. The truncated normal distribution is specified by its mean $\mu_{\tau_0}$, variance $\sigma^2_{\tau_0}$, and truncation limits $t'_0$ and $t'_m$ such that $t'_0 < \mu_{\tau_0} < t'_m$. These hyperparameters are chosen to reflect prior knowledge about the plausible range of the initial age shift, while remaining relatively uninformative within those bounds..

\subsubsection{Prior for $\sigma$}

For the single and double target alignment strategies, where the standard deviation $\sigma$ of the t--distribution is unknown, we assign a Gamma prior distribution:
\begin{equation}
\sigma \sim \text{Gamma}(\alpha_\sigma, \beta_\sigma)
\end{equation}
The Gamma distribution is a common choice for scale parameters and ensures that $\sigma$ remains positive. The shape ($\alpha_\sigma$) and scale ($\beta_\sigma$) hyperparameters can be set based on prior knowledge or expert opinion about the expected range and distribution of the standard deviation. By default, we set these parameters to $1.5$ and $.01$, respectively. This allows prioritizing the reduction of this parameter, yielding a prior mean of 0.01 and a 95\% quantile of approximately 0.03. However, since the parameter follows a gamma distribution, its value can be larger if required. These priors are weakly informative, allowing flexible estimation of $\sigma$ without overly constraining it.

\subsubsection{Prior for $\omega_m$}

Here, $\omega_m$ is a mixing parameter that determines the relative contribution of each target record $v^1$ and $v^2$ in the combined target signal $v$. We assign a uniform prior distribution to $\omega_m$, effectively treating both target records as equally important \textit{a priori}:
\begin{equation}
\omega_m \sim \textrm{Uniform}(0, 1) \quad \textrm{or equivalently} \quad \omega_m \sim \textrm{Beta}(1, 1)
\end{equation}
The uniform/Beta distribution allows $\omega_m$ to take any value between 0 and 1. This enables the model to infer which target record ($v^1$ or $v^2$) has a stronger influence on the alignment, i.e. the posterior distribution of $\omega_m$ reveals which target record aligns more closely with the input, adapting dynamically based on the observed information. This choice reflects the \textit{a priori} belief that neither target is inherently more suitable; users can adjust the Beta distribution's shape parameter to place greater prior weight on a given record. It is important to note that, this weighting is treated as a global parameter.

\subsubsection{Prior for $\tau_n$}

To facilitate the alignment process and enable users to place the input record within a specific time window, we allow users to specify a prior for the age of the last input data point. This prior is modeled as a normal distribution with a mean of $\tau_n$ and a standard deviation of $\sigma_t$, both of which can be set by the user. Notably, since $\tau_n \in T'$, this prior is represented by a truncated normal distribution, expressed as:
\begin{equation}
\tau_n \sim \text{TruncNormal}(t_n, \sigma_{t_n}, t'_0, t'_m).
\end{equation}
This prior enables the method to focus the search for alignment within a plausible age range, thereby eliminating unnecessary alignments and enhancing the efficiency and precision of the alignment process. The flexibility in defining $t_n$ and $\sigma_{t_n}$ also allows users to incorporate prior knowledge, thus aligning the model more closely with the specific characteristics of their data.

\subsection{Computational implementation}

To generate samples from the posterior distribution and efficiently explore the parameter space, we employ a MCMC methodology using the t--walk algorithm \citep{christen2010}. This algorithm is used by other commonly used age--depth modeling routines, such as \texttt{Bacon} \citep{Blaauw2011}, which provides additional guidelines for its application. Notably, \citet{Blaauw2011} suggest that convergence is typically achieved by $100 \times n \times \text{thinning}$, where $n$ represents the number of parameters in the model. They also recommend saving every $n \times \text{thinning}$ iteration to ensure a pseudo--independent sample. With these settings, we achieve an integrated autocorrelation time (IAC) of less than 50 MCMC iterations in most cases.

These suggested settings provide a platform with minimal user interaction, which is beneficial, as the method is designed to be practical for users across different levels of statistical training. The time required to obtain 3,000 effective samples depends on the sample size of the input record but typically takes 1--2 hours to complete the MCMC for a model with about 50 parameters. These tests were conducted on an Apple M1 MacBook Pro. This time significantly increases if we opt for the model that incorporates age uncertainty from the target, as the calculation of the kernel distribution is computationally intensive.

The implementation was developed in \textit{R}, with the likelihood and t--walk functions optimized using \texttt{Rcpp} \citep{Eddelbuettel_2011} to enhance computational efficiency. The software release is available in \citep{aquino_bsync_2025}. A GitHub repository is also available at \href{https://github.com/maquinolopez/BSynch}{https://github.com/maquinolopez/BSynch}.

\section{Alignment of environmental records}\label{sec:classical_ali} 

In this section, we demonstrate the capabilities of our single target methodologies using a suite of records from the Iberian Margin. This region is an ideal case study as it provides numerous widely used records capturing well--defined millennial--scale variability \citep{Skinner_2019, Waelbroeck_2019, Shackleton_2000}, enabling comparisons of various alignment and age--modeling strategies. First, we introduce a classical alignment strategy by aligning the benthic $\delta^{18}$O record from core MD95--2042 \citep{Schonfeld_2003, Lisiecki_2022} to  \emph{Prob--stack} \citep{Ahn_2017} --a probabilistically stacked compilation of nearly 200 benthic $\delta^{18}$O records. It should be noted that the \emph{Prob--stack} was constructed using using MD95--2042. Although this introduces some circularity, the contribution of MD95--2042 to the overall stack is minor, and this example is intended primarily for heuristic purposes. Second, to illustrate the flexibility of \texttt{BSync} in synchronizing different types of data sets, we align a planktic $\delta^{18}$O records from core MD03--2698 \citep{Lebreiro_2009} to its counterpart in core MD95--2042 under two scenarios: one assuming no age uncertainty in the target and another incorporating the target's reported age--model uncertainty.  Because the double target represents a novel feature that has not been previously tested, we evaluate it through a controlled experiment using synthetic data to provide a clearer demonstration of its functionality in Section \ref{sec:comp_exp}.

\subsection{Alignment of Benthic Records}

\begin{figure}[htbp]
\centering
\includegraphics[width=\textwidth]{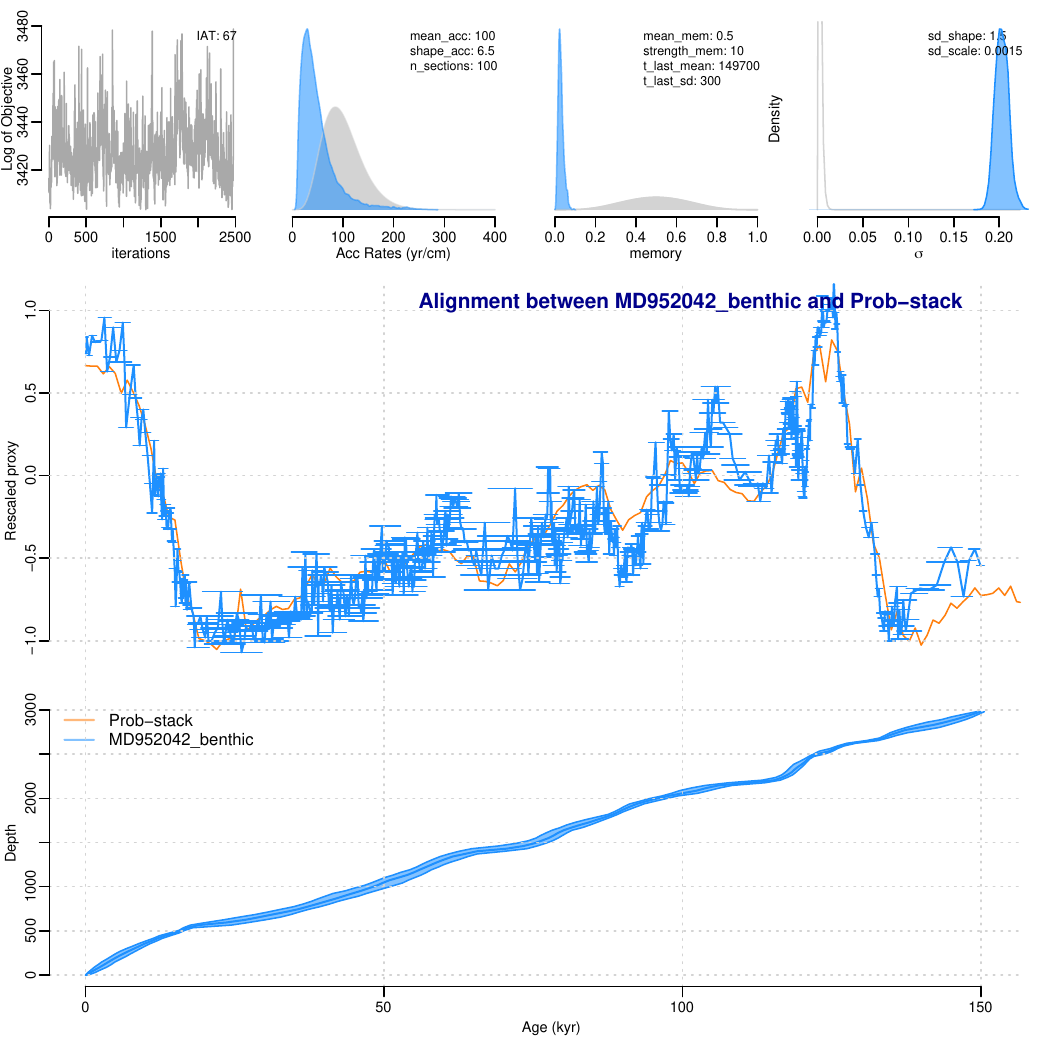}
\caption{\texttt{BSync}'s alignment between MD95--2042 benthic  $\delta^{18}$O record and \emph{Prob--stack}, shown in the style of default plots generated by  \texttt{BSync}.
Top row (left to right): Log--objective trace of the MCMC, as commonly used in methods like \texttt{Bacon} \cite{Blaauw2011} to check for convergence by visually inspecting for the absence of step--like sections; prior (grey) and posterior (blue) distribution of deposition time parameters (yr/cm); prior (grey) and posterior (blue)  distribution of the memory parameter ($\omega$) and the standard deviation of the alignment ($\sigma$).
Second row: Proxy records for the target (\emph{Prob--stack}, mean uncertainty $1\sigma \approx 0.18 \permil$) and the aligned input (MD95--2042) (blue).
Third row: Age--depth model (or age--age model for age--scale correction) with 95\% credible intervals, alongside the originally reported age model.}
\label{fig:benthic_alignment}
\end{figure}

We apply \texttt{BSync} to align the MD95--2042 benthic $\delta^{18}$O record \citep{Schonfeld_2003, Lisiecki_2022} to the \emph{Prob--stack} \citep{Ahn_2017}. Figure \ref{fig:benthic_alignment} shows the alignment results formatted to replicate the standard output generated by \texttt{BSync}, which includes model diagnostics (top panel), proxy alignment (middle panel), and the resulting age--depth model (bottom).   The model was run with 100 equally spaced sections in depth (which we denote as \texttt{Bacon} sections) and a prior mean deposition time of 100 years per centimeter to accommodate a relatively wide range of deposition times. The prior on the memory parameter was set to default values to allow \texttt{BSync} learning from the data. 

The top panel of Figure \ref{fig:benthic_alignment} shows the log--likelihood trace, suggesting stable MCMC sampling, whereas the prior vs. posterior distributions (grey vs. blue) for the model deposition time, memory parameter, and alignment uncertainty ($\sigma$) illustrate how \texttt{BSync} updates its beliefs. In this case, the contrast between prior and posterior distributions, particularly for deposition times, highlights that \texttt{BSync} not only relies on prior information for alignment but also actively learns from the data to achieve an optimal alignment. 

The middle panel of Figure \ref{fig:benthic_alignment} shows both the target (\emph{Prob--stack}) and aligned input (MD95--2042) proxy values, highlighting a good structural match between the records. The bottom panel displays the resulting age--depth relationship of the aligned MD95--2042 data, and its 95\% credible intervals. The posterior age model for MD95--2042 does not contain large fluctuations in sedimentation rate. This is broadly consistent with the expectation of a relatively smooth depositional history and realistic age--depth uncertainties, as well as the smoothed nature of benthic $\delta^{18}$O signals, which track long--term changes in global ice volume and deep--water temperatures.

These results demonstrate that \texttt{BSync} can efficiently align benthic $\delta^{18}$O records while allowing for a smooth but flexible age--depth relationship and modest variability in the proxy signals. In the following section we consider a more challenging scenario: the alignment of noisier planktic proxy data.

\subsection{Alignment of planktic proxy data}

Planktic $\delta^{18}$O records generally feature greater high--frequency variability (''noise'') than benthic records because they are influenced by rapidly changing surface--water conditions. This higher variability requires caution when aligning records, as not all short--term features are spatially coherent or represent regional climate \citep{Lee_2023}. Nevertheless, planktic $\delta^{18}$O records provide valuable insights into past ocean conditions, including surface temperature and salinity \citep{Shackleton_1987}. In this example, we use the MD03--2698 record \citep{Gouzy_2004, Lisiecki_2022} as our input, and the MD95--2042 record --recovered less than 100 km away-- as our target. Given that both records represent $\delta^{18}$O measurements from the same species and from nearby locations, we expect them to integrate coherent oceanographic signals.

To evaluate \texttt{BSync}'s performance under the standard single--target alignment case, we conducted two tests.  The first test aligned the MD03--2698 record to MD95--2042 under the assumption of no age--uncertainty. The second instead factored in age--uncertainty by constructing a Bayesian age--depth model for MD95--2042 using \texttt{Bacon} \citep{Blaauw2011}, which generates iterations from the posterior distribution of the age scale. These iterations were incorporated into our methodology using the kernel--based approach described in Section \ref{sec:Likelihood} (see also \citet{Aquino_2024}). In both tests we split the input record into 50 \texttt{Bacon} sections with prior deposition times obtained from the reported age--depth model (see Section \ref{sec:accu_m}). The total number of iterations (85,800,000) was determined by the desired 3,000 retained posterior samples, given a burn--in of 30,000 iterations and a thining factor of 25,000.

\begin{figure}[htbp]
	\centering
	\includegraphics[width=\textwidth]{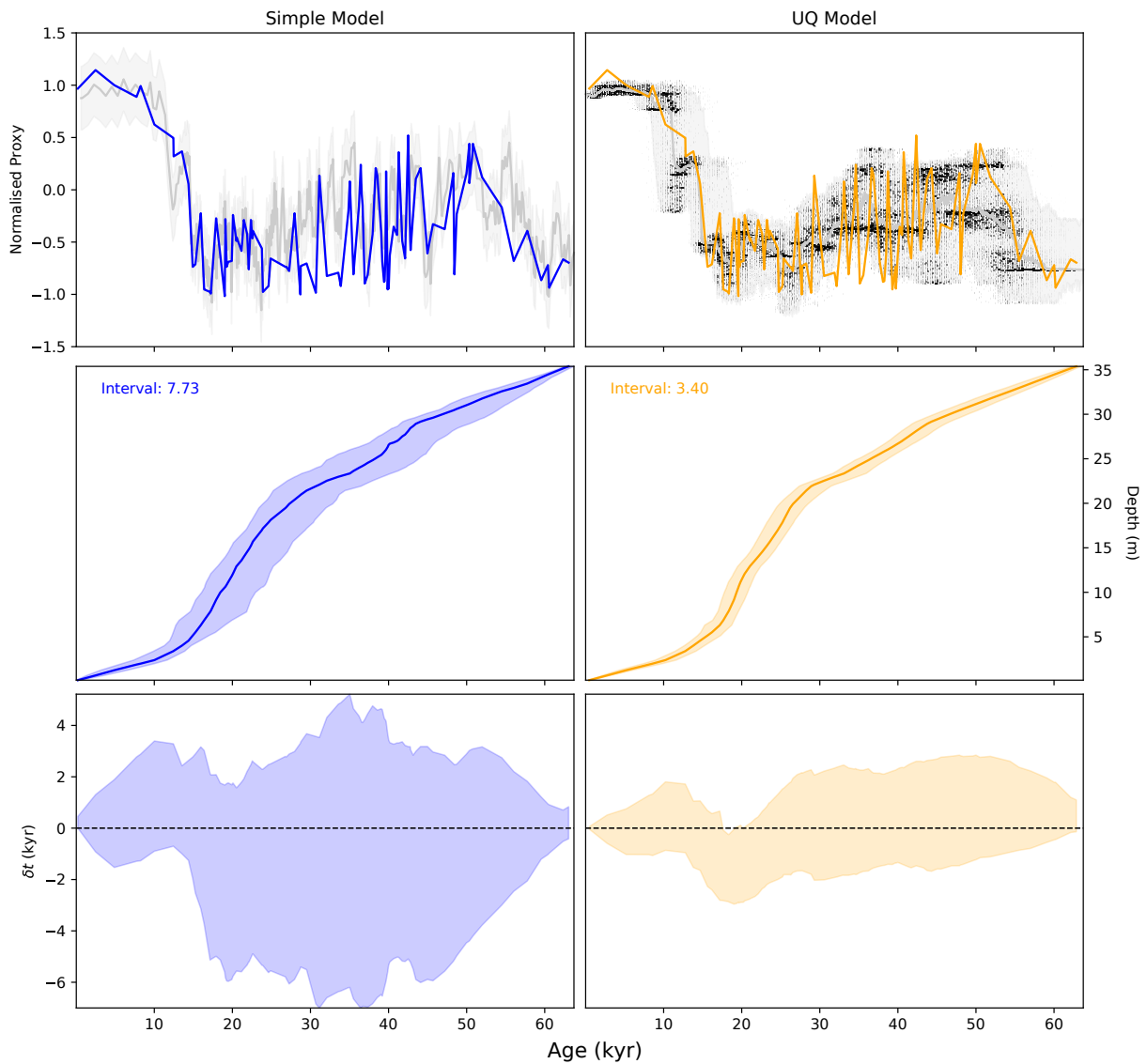}
  	\caption{
    Synchronization of planktic $\delta^{18}$O records MD03--2698 to MD95--2042 using  \texttt{BSync}. Each column corresponds to a different strategy: single target (``Simple Model"; blue, left), and single target with age uncertainty quantification (``UQ Model"; yellow, right).
    	Top row: Alignment of the proxy record (colored lines) to the corresponding target (gray line).
    	Middle row: Optimal age--depth models produced by  \texttt{BSync} (colored lines) with associated uncertainty bands (95\% credible interval, shading).
    	Bottom row: Differences ($\delta_t$) between the originally reported age--depth scale and the age--depth scale obtained by  \texttt{BSync} as a function of time.
            }
	\label{fig:combined}
\end{figure}

Figure \ref{fig:combined} summarizes the results of the single--target alignments. The top panel displays the input proxy signals on their median age scale following alignment. The blue signal represents the standard single--target strategy, while the yellow signal corresponds to the strategy that incorporates the age uncertainty of the target record. For reference, the gray transparent envelope depicts the target record (MD95--2042) with two standard deviations based on the mean posterior of the $\sigma$ parameter in the single--target alignment model; the darker intervals indicate regions of higher data density, while the shadings represent the 95\% quantiles.  The middle panel shows the resulting age--depth relationships obtained by aligning the records. Finally, the bottom panel displays the offset ($\delta t$) between the radiocarbon--based age scales and those obtained via our alignment, along with their corresponding 95\% credible intervals.

In the case of the standard single--target strategy, the rescaled target signal is bracketed by uncertainty bands of approximately 0.6 normalized units. Since the proxy signal is rescaled to fit within the range $[0,1]$, this reflects 30\% to 60\% of the entire proxy scale. This conservative estimation of the standard deviation of the proxy signal accounts for the limitation of having no age uncertainty in the alignment target. Given the assumed lack of age errors, we deem the resulting uncertainty to be plausible.  

When the age model uncertainty is included in the alignment, this error propagates into posterior uncertainty in the target proxy signal \citep{Aquino2020}  (top panel of Figure). Unlike the previous strategy, the resulting uncertainty is not centered on a single value; rather, the posterior concentrates in regions where the target data are most dense, producing bands of higher probability. From Figure \ref{fig:combined}, we observe that incorporating age uncertainty reduces the overall alignment error by up to 60--80\% relative to the alignment with no age uncertainty.

Overall, these tests show that both strategies provide sensible alignments with age uncertainties consistent with the available information provided to the model. Both strategies yield posterior distributions with relatively narrower age errors where there is sufficient structure for aliment between the input and target records. This pattern is generally expected, as the resulting uncertainties are consistent with the information available to the model.

%%%%%%%%%%%%%%%
\section{Simulation study}\label{sec:comp_exp}

As the true sedimentation history of paleoenvironmental records is inherently uncertain \citep{Blaauw2018}, we test the performance of our model using synthetic simulations. We define a simulated age--depth function $t(x)$, as the ground--truth target for \texttt{BSync}. Specifically \texttt{BSync} seeks $\tau(x)$ that is as similar as possible to the true age--depth function, i.e., $\tau(x)\rightarrow t(x)~\forall~x\in D$). We generate $t(x)$ by specifying an deposition time at depth $x$,
\begin{equation}
dt(x) = \frac{d}{b} \cdot x \left(0.9 - \cos\left(\pi \cdot \frac{x}{b}\right)\right) + a,
\end{equation}

\noindent
where $dt(x) > 0$ for all $x$ ensures a continuous, monotonically increasing age--depth function.  Here, $a = 20~yr~cm^{-2}$ denotes the deposition time at the surface, $b = 250$ represents the half cycle periodicity, and $d = 12 $ controls the amplitude of variation. Integrating $dt(x)$ with $t(0) = 0$ gives:

\begin{equation}\label{eq:t_simu}
	t(x) = \frac{{d \cdot b}}{{\pi^2}} + a \cdot x + \frac{{d \cdot 1.5 \cdot x^2}}{{2 \cdot b}} - \frac{{d \cdot x \cdot \sin\left(\frac{{\pi \cdot x}}{{b}}\right)}}{{\pi}} - \frac{{d \cdot b \cdot \cos\left(\frac{{\pi \cdot x}}{{b}}\right)}}{{\pi^2}} ,
\end{equation}

Figure \ref{fig:combined_plots} illustrates $t(x)$ (Panel a) and its derivative $dt(x)$ (Panel b) over the domain $x \in [0, 1000]$ cm, where $t(x)$ shows a continuous age--depth relationship with a linear trend modulated by periodic fluctuations, and $dt(x)$ reflects alternating periods of relatively faster and slower sedimentation.

We here emulate a sediment--like record with an unknown true age model by constructing an artificial proxy series from two well--dated target $\delta^{18}$O ice--core records: NGRIP from Greenland ($v_g$) \citep{NGRIP2004} and EPICA Dome C (EDC) from Antarctica ($v_a$; \citep{EPICA2004}). Both records have been recently placed on the AICC2023 timescale \citep{Bouchet2023AICC2023}. We generate a composite climatic signal as a weighted mixture of the two records, $p(t) = 0.7 v_g(t) + 0.3 v_a(t)$, representing a mixed signal arising from both sites. We then convert this time series into a sediment record by warping it with the simulated age--depth function, $t(x)$. It should be borne in mind, that the purpose of this exercise is not to create a synthetic ice--core record, but rather an idealized sediment target record that integrates ice--core climate signals in different proportions. Finally, this warped signal is interpolated to fixed depth intervals in order to replicate practical sampling/discretization errors, with each proxy data point thereafter perturbed with randomized noise, $p_i \sim \mathcal{N}(C_1,p(t(d_i)) + C_2,, \delta,\sigma)$, to simulate measurement error and depositional noise.  Here, $C_1$ and $C_2$ are constants, $\sigma$ is the mean residual standard deviation from a Loess regression \citep{Cleveland_1992}, and $\delta$ is set to 0.05 (i.e., 5\% added noise; note that similar alignment results are found up to 30\% of added noise, Figure \ref{fig:simulation_results}). This framework therefore generates a noisy synthetic record with a perturbed (but known) age scale, allowing us to test whether \texttt{BSync} can recover \emph{both} the original age--depth relationship and the mixing proportion when aligning the record back to the original ice core targets.

%%%%%%%%%%%%%%%
\subsection{Simulation study results}

\begin{figure}[htbp]
	\centering
	\includegraphics[width=1\textwidth]{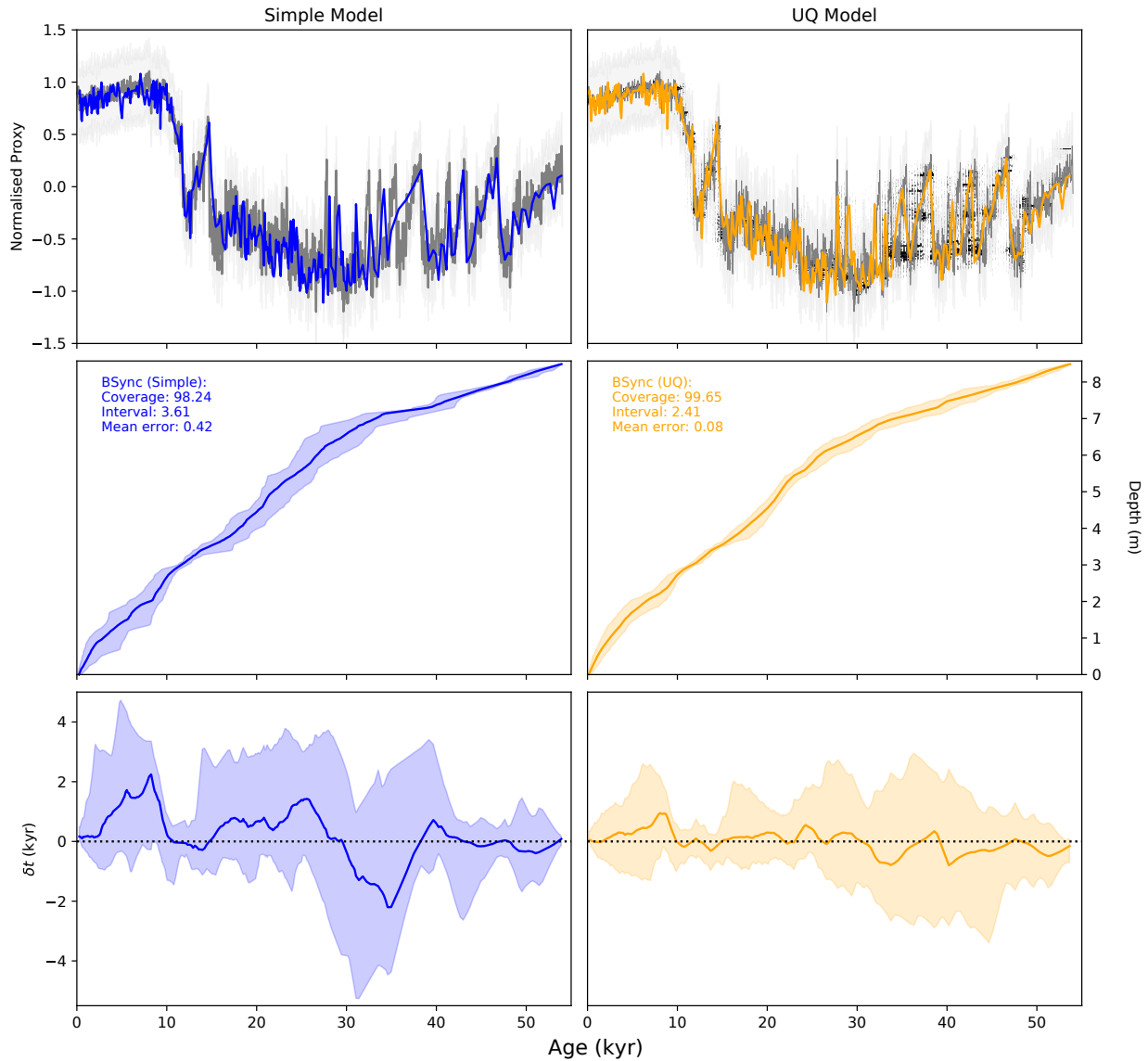}
  	\caption{\texttt{BSync}'s alignment between the simulated input record and NGRIP. Each column corresponds to a different strategy: (left) single target and (right) single target with age uncertainty.
    	Top row: Alignment of the proxy record (colored lines) to the corresponding target(s) (gray line).
    	Middle row: Age--depth models produced by  \texttt{BSync} (colored lines) with associated uncertainty bands (95\% credible interval).
    	Bottom row: Deviations ($\delta_t$) from the true age expressed in units of standard deviations.
        }
	\label{fig:simulation_results}
\end{figure}

The results shown in Figure \ref{fig:simulation_results}, illustrate that \texttt{BSync} accurately recovers the true age--depth function across different alignment scenarios. In both the single--target and UQ strategies, the reconstructed age models closely track the true chronology, with the 95\% credible intervals encompassing the true age--depth function. Notably, the single--target alignment shows a broader uncertainty band (average interval width of 3.61 kyr) and a higher mean error ($\delta_t$) of 0.42 kyr, whereas the UQ model consistently yields a narrower credible interval (2.41 kyr on average) and a smaller mean error of 0.08 kyr. 
It is worth noting that the single--target strategy partly fails to align the Dansgaard--Oeschger millennial variability between 30--40 kyr, whereas the UQ approach provides a more sensible synchronization. This is likely a result of the additional information incorporated into the UQ strategy.
Nonetheless, the top panel shows how both aligned proxies (colored lines) closely follow the target signal (gray), especially in intervals with clear data structure. Together, these results highlight \texttt{BSync}'s ability to adaptively adjust uncertainty in alignment scenarios involving a double target and to produce robust age--depth models under unknown target mixtures.

\begin{figure}[htbp]
	\centering
	\includegraphics[width=0.95\textwidth ]{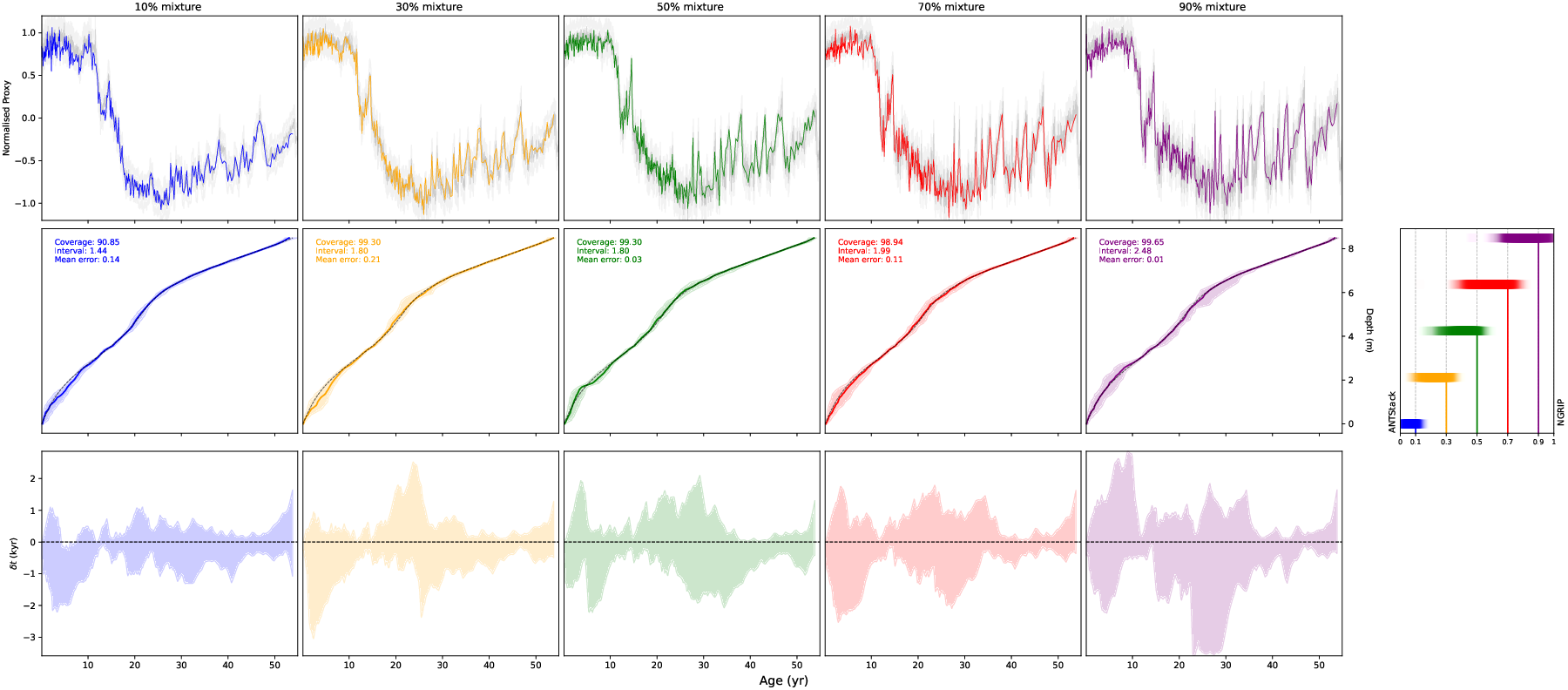}
    \caption{
        BSync's alignment results for the double--target approach. The alignment of a synthetic record, created as a mixture of NGRIP \citep{NGRIP2004} and Antarctic Stack (ANTStack) \citep{AntarcticStack}, to a true target record. Each column represents a different mixture ratio, where the percentage indicates the proportion of NGRIP variability in the synthetic record (e.g., ``95\% mixture" corresponds to 95\% NGRIP and 5\% ANTStack).\\
        The top panels show the alignment of the synthetic input record to the true target record, the middle panels show the resulting age--depth relationship, and the bottom panels show the deviations from the true age expressed in units of standard deviations. The true age--depth relationship is shown in black. The right panel shows the posterior distributions of the mixing parameter for each alignment, with colors corresponding to the mixture proportions.
    }
	\label{fig:mixing_results}
\end{figure}

Building on the above synthetic test, we next evaluate how well the double--target approach can recover differing mixture proportions between two target records.  Figure~\ref{fig:mixing_results} shows the results from this test for scenarios with the NGRIP contribution gradually increasing from 10\% to 90\%. Across all scenarios, \texttt{BSync} successfully recovers both the true age--depth function and the true mixing parameter with posterior distributions closely centered around the pre--assigned mixture, as shown in the right--most panel. The top row displays the aligned proxy signals (colored) relative to the target (gray), demonstrating excellent synchronization performance regardless of the mixing ratio. The middle row shows the resulting age--depth models with 95\% credible intervals, which consistently contain the true chronology. Notably, the coverage (i.e. the proportion of the credible intervals which contain the true value) exceeds 98.9\% for all but the 10\% NGRIP case, where it remains high at 90.85\%. The mean error decreases steadily as NGRIP's contribution increases (from 0.14 kyr at 10\% to 0.01 kyr at 90\%), while interval widths remain relatively narrow throughout (1.44--2.48 kyr).

%\begin{landscape}
%\end{landscape}

Altogether, these synthetic tests demonstrate that \texttt{BSync} effectively and efficiently recovers the true age--depth relationship and target record mixture proportions. 
Although the choice of prior can influence the outcome, the tests indicate that, as long as priors are not overly restrictive, the model consistently converges to accurate age--depth reconstructions. These findings support the use of \texttt{BSync} in paleoclimate applications where an input record must be aligned to, or compared against, multiple target records\citep{Rohling2009}.

%%%%%%%%%%%%%%%%%%%%%%%%%%%%%%%%%%%%%%%%
%%%%%%%%%%%%%%%%%%%%%%%%%%%%%%%%%%%%%%%%
\section{Comparison with BIGMACS}\label{sec:bigmacs}

Both \texttt{BSync} and \texttt{BIGMACS} \citep{Lee2023} tackle the challenge of aligning proxy time series to place climate records on a common age scale with minimal user input. Since \texttt{BIGMACS} is, to the best of our knowledge, the most established tool for automated alignment, it provides a useful benchmark against which to evaluate the performance of our routine.
Methodologically, however, they differ. \texttt{BIGMACS} models the alignment transformation via a Gaussian process and relies on fixed, empirical priors. In contrast, \texttt{BSync} employs an inverse--gamma process with flexible, interpretable priors that can be adapted to the specifics of a given depositional scenario. Another key distinction lies in the likelihood treatment: \texttt{BIGMACS} requires the user to specify proxy uncertainty; by contrast, \texttt{BSync} either (i) infers proxy uncertainty jointly with the synchronization, or (ii) takes uncertainty from an external Bayesian age--depth model and propagates it into the target signal. To evaluate the sensitivity of these design choices, we present a set of simulations where we compare the two routines under identical conditions. In particular, we examine how each routine performs in the presence or absence of age constraints, with special focus on the impact of age constraints. A full benchmarking experiments is provided in Appendix~\ref{appendix:additional_test}.

We revisit the simulation described in Section~\ref{sec:comp_exp} to examine how \texttt{BSync} and \texttt{BIGMACS} perform under varying levels of proxy uncertainty and the presence or absence of age constraints. In this experiment, both methods are applied to the same synthetic record with a known age--depth relationship. 
While \texttt{BIGMACS} requires an explicit age constraints to infer the youngest and oldest ages, \texttt{BSync} can, in principle, use the prior on $\tau_0$ and $\tau_n$ to play this role. Note that for this particular test, no age constraint is provided at the bottom of the core ($\tau_n$) for either method.
This allows us to assess two important questions: (1) how sensitive is \texttt{BIGMACS} to the level of proxy uncertainty, and (2) how do both methods perform when alignment is inferred with minimal chronological information? This approach thus isolates the influence of internal model structure, particularly the performance of the link function of each model.

\begin{figure}[htbp]
	\centering
	\includegraphics[width=1\textwidth]{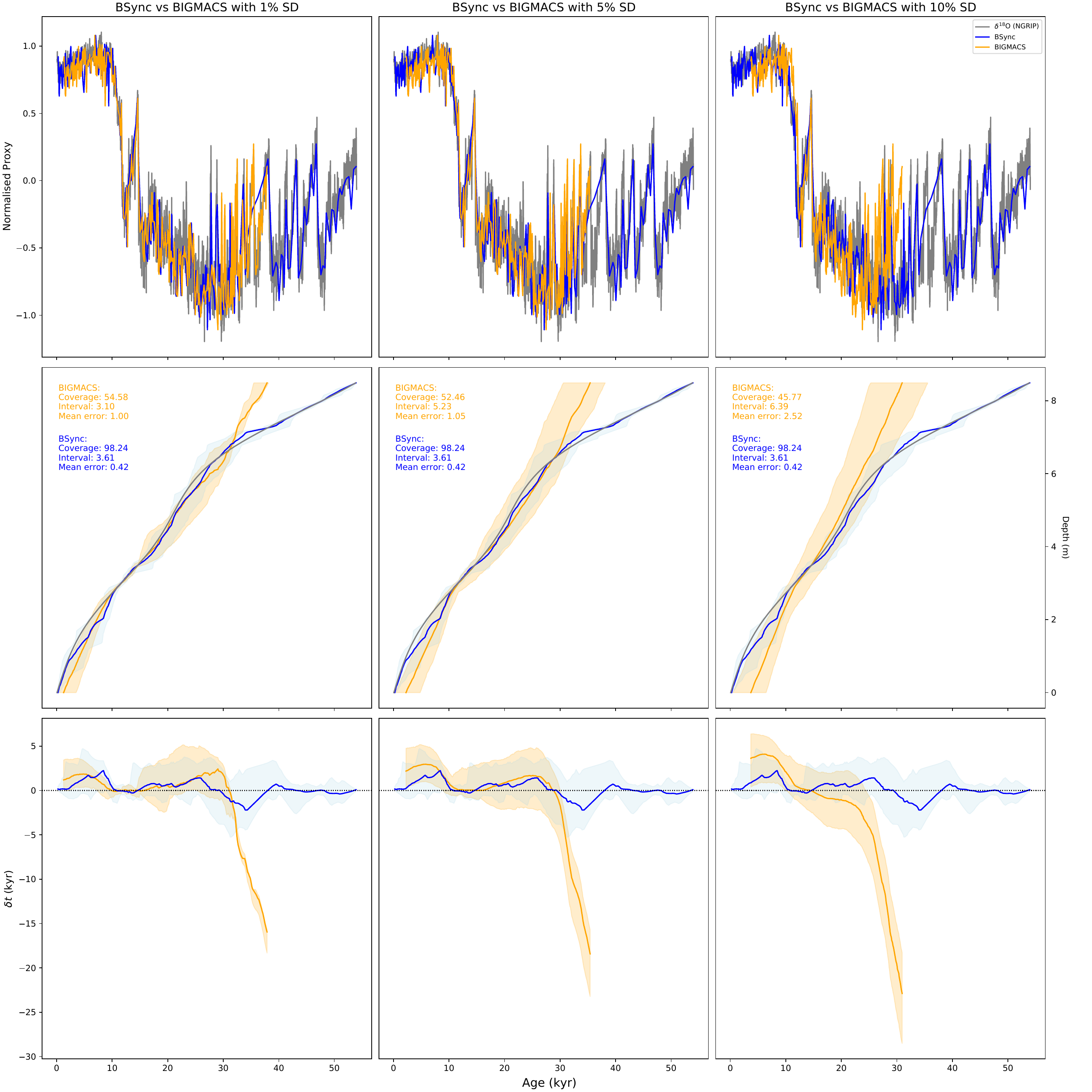}
  	\caption{Comparison of \texttt{BIGMACS} (orange) and  \texttt{BSync} (blue) using the simulated record described in section \ref{sec:comp_exp}. For BIGMACS, results are shown for different assumed levels of target proxy uncertainty. The top panels show proxy synchronizations, the middle panels show the resulting age--depth relationship, and the bottom panels show the deviations from the true age expressed in units of standard deviations. The true age--depth relationship is shown in black.}
	\label{fig:Bigmacs_sd}
\end{figure}

We tested both questions using three levels of proxy uncertainty in \texttt{BIGMACS}: 1\%, 5\%, and 10\% noise applied to the scaled target record. As shown in Figure~\ref{fig:Bigmacs_sd}, the alignment output is highly sensitive to proxy noise: as uncertainty increases across the three experiments, the average credible interval widens from 3.1 kyr to 6 kyr, while the mean error rises from 1 kyr to 2.5 kyr. Among these, the 1\% uncertainty case yields the closest results to those from \texttt{BSync}, which produced --under the same simulated conditions-- an average interval length of 3.61 kyr and a mean error of 0.42 kyr.

In terms of coverage, the most striking difference is that \texttt{BIGMACS} consistently reports values near 50\%, which means that about 50\% of the true ages lie outside of the reported intervals, regardless of the applied level of proxy uncertainty. This indicates that, in the absence of age constraints, BIGMACS fails to provide an accurate synchronization and to quantify alignment uncertainty reliably. Further, inspection of the results (Figure~\ref{fig:Bigmacs_sd}) reveals that \texttt{BIGMACS} often defaults to an approximately linear age--depth relationship, largely ignoring the structural features of the input signal. This behavior is likely a consequence of the lack of chronological tie points --particularly the missing constraint on $\tau_n$-- which deprives the model of the information needed to nudge the alignment toward the true chronology. Conversely, when age constraints are absent, \texttt{BSync} yields a significantly higher coverage (98.24\%), with credible intervals that vary in parallel with the complexity of the aligned signal, and more accurately reflect posterior uncertainty.

To determine whether these results could be explained by low sampling resolution or high measurement noise, we conducted additional sensitivity tests, outlined in the Appendix~\ref{appendix:additional_test}). The results show that \texttt{BIGMACS} consistently under--performs across a broad range of sampling resolution and noise levels, suggesting that its reliance on age constraints is likely the main limitation of this approach. BSync, instead, consistently yields accurate and comparable results across these sensitivity experiments, even when the data exhibit low resolution and/or high levels of noise. 

However, since additional age constraints are often available, it is also important to assess how both routines perform when such information is provided. To explore this, we repeated the simulation experiment described above by gradually adding age constraints --i.e. 2, 4, 6, 8, and 10-- distributed randomly along the stratigraphy, in addition to the fixed constraints at the top and the bottom. Each constraint was simulated as a normal random variable centered at the true age, with a standard deviation of $\sigma = 0.25$ kyr.

While both routines can incorporate chronological information, they do so in fundamentally different ways. \texttt{BIGMACS} integrates the age constraints directly into the alignment process, treating them as fixed inputs. \texttt{BSync}, on the other hand, assumes that a Bayesian age--depth model has already been constructed from these constraints, and uses the resulting posterior distribution to inform \texttt{BSync}'s priors. Here, we fit a \texttt{Bacon} model using the simulated age estimates, and used its posterior to define priors for \texttt{BSync}, following the procedure described in Section~\ref{sec:accu_m}.

%\begin{landscape}
\begin{figure}[htbp]
	\centering
	\includegraphics[width=0.95\textwidth]{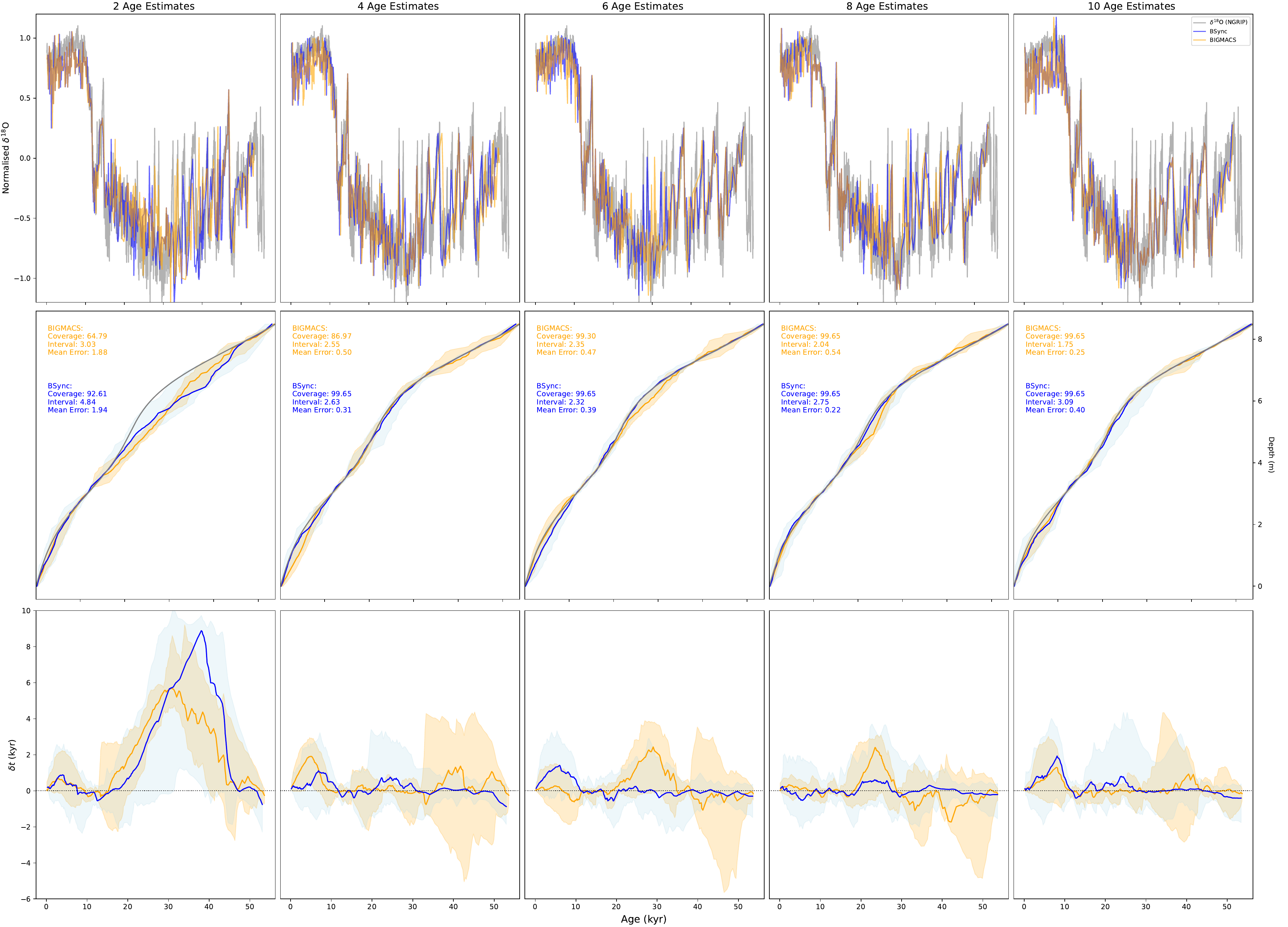}
  	\caption{Comparison of \texttt{BIGMACS} (orange) and  \texttt{BSync} (blue) using different numbers of age constraints (left to right). The top panels show proxy synchronizations, the middle panels show the resulting age--depth relationship, and the bottom panels show the deviations from the true age expressed in units of standard deviations. The true age--depth relationship is shown in black.} 
	\label{fig:8}
\end{figure}
%\end{landscape}

Figure~\ref{fig:8} shows the results of these experiments. As anticipated, \texttt{BIGMACS} performs systematically better as more age constraints become available. Its weakest performance occurs when only two age constraints are present, yielding a coverage of 64.8\%. This is the most challenging condition for both methods; nevertheless, \texttt{BSync} achieves substantially higher coverage (92.61\%) than \texttt{BIGMACS}. Interestingly, \texttt{BSync} maintains coverage values above 90\% across all scenarios. In the remaining experiments, coverage improves notably for \texttt{BIGMACS}, exceeding 90\% in all cases except the one with four age constraints. In terms of accuracy, the scenario with two constraints also yields the largest mean errors, whereby \texttt{BIGMACS} reaches 1.88 kyr and \texttt{BSync}  1.94 kyr. Mean errors decrease substantially in the other experiments: \texttt{BSync} yields errors in the range of 0.22--0.40 kyr as additional constraints are introduced. Finally, we evaluate interval lengths, using this metric to assess how each method improves as additional age constraints are incorporated. In the most weakly constrained setting, \texttt{BIGMACS} yields a mean interval length of 3.03 kyr, while \texttt{BSync} produces wider intervals averaging 4.84 kyr. As more age constraints become available, \texttt{BIGMACS} exhibits a nearly linear reduction in interval length, decreasing by approximately 0.3 kyr for every two additional constraints. By contrast, \texttt{BSync} does not show a systematic reduction in interval length; instead, its intervals remain relatively stable, ranging between 2.3 and 3.0 kyr. Overall, these results suggest that \texttt{BIGMACS} performs well when sufficient chronological information is available, but its accuracy deteriorates when age constraints are sparse. On the other hand, \texttt{BSync} remains consistently reliable in quantifying uncertainty, even if its intervals are slightly wider on average.

%%%%%%%%%%%%%%%
\section{Conclusion and discussion}\label{sec:discussion}

Here we present \texttt{BSync}, a new Bayesian model for synchronizing two proxy time series and generating age scales for paleoclimate records. \texttt{BSync} advances existing alignment approaches by formulating time series synchronization as a Bayesian inference problem, yielding interpretable and flexible age--scale posteriors.

Traditional proxy alignment methods often struggle to balance interpretability, uncertainty quantification, and adaptability across diverse proxy types, climate archives, or depositional environments.  Our model helps to overcome these limitations in several ways. First, \texttt{BSync} provides a flexible parametric family of alignments based on a generalized inverse--gamma process, which accommodates physically realistic sedimentation--rate variability and supports interpretable prior specifications that can be easily adjusted on a per--site basis. Second, it employs a robust likelihood formulation, using t--distributions and/or kernel--based uncertainty modeling, to account for the noisy nature of proxy data in paleoclimate archives. Third, \texttt{BSync} extends existing frameworks by enabling alignment against either single or mixed targets, and by providing a proper treatment of reported age uncertainty in target records.

\texttt{BSync} further contributes to tackling the broader problem of uncertainty propagation across successive stages of Bayesian analysis. Specifically, we demonstrate two distinct strategies for incorporating information from prior Bayesian analyses into \texttt{BSync}. The first approach transforms the posterior information from a previous Bayesian age model into prior distributions for the parameters of the alignment model, thus preserving and formalizing prior knowledge in the subsequent stage (Section \ref{sec:accu_m}). The second strategy treats the posterior MCMC samples from the previous model as the likelihood of the process. This is achieved via kernel density estimation, thereby allowing the full uncertainty to be incorporated into the new inference (Section \ref{sec:Likelihood}). These two strategies provide a flexible approach for uncertainty--aware model composition.

In a suite of sensitivity tests, we demonstrate that \texttt{BSync} achieves well--calibrated uncertainty intervals, with coverage consistently exceeding 90\% even in the most challenging alignment scenarios. The model successfully aligns structures in the proxy data and adapts effectively to varying levels of noise, sampling resolution, availability of prior information, and number of age constrains. In comparative experiments, \texttt{BSync} outperforms existing routines (BIGMACS) in both accuracy and precision, particularly when age constraints are scarce or unavailable.

%More broadly, \texttt{BSync} shows how Bayesian modeling provides a natural and effective framework for addressing time series synchronization problems for geophysical data. The strategies we propose for integrating information across sequential Bayesian stages also contribute to the growing literature on uncertainty propagation in complex scientific workflows.

Future work will focus on improving the computational efficiency of \texttt{BSync}, through optimized implementation in a compiled language. We will also explore more refined approaches for modeling prior information from Bayesian age--depth models and for extending the framework to joint alignment of multiple proxy records, i.e mixing multiple targets. We see considerable potential for integrating \texttt{BSync} with existing Bayesian tools for paleoclimate studies to support fully end--to--end uncertainty quantification in paleoclimate workflows.

The R implementation is openly available at \href{https://github.com/maquinolopez/BSynch}{https://github.com/maquinolopez/BSynch}.

%\section{Bibliography}
%\bibliographystyle{apalike}
\bibliographystyle{agsm}
\bibliography{bibliography}  % Adjust to match your .bib filename

\section*{Data and Software Availability}

The BSync software developed for this study is available as an open--source \textit{R} package at \url{https://github.com/maquinolopez/BSynch}. A snapshot of the software and data used in the analysis is archived at Zenodo with DOI \url{https://doi.org/10.5281/zenodo.15785856}.

\section*{Conflict of Interest}

The authors declare no conflicts of interest.

\section*{Acknowledgements}
This research has been supported by the Isaac Newton Trust (grant no. LCAG/444.G101121) and a Natural Environment Research Council (NERC) Discovery Science Grant (NE/W006243/1) awarded to Francesco Muschitiello. This study is a contribution to the INTIMATE (INTegration of Ice--core, Marine, and Terrestrial records) project.

\section*{Author Contribution}

MAL led the conceptualization of the study, developed the mathematical framework and formalized the ideas, designed the methodology, implemented the software, designed and conducted the experiments, performed the formal analysis, and wrote the original draft of the manuscript. FM conceived the project and secured funding, contributed to the conceptual development and methodology, participated in the evaluation and interpretation of the results, and contributed to writing, reviewing, and editing the manuscript. MO contributed to conceptualization, testing and evaluation of the results, and writing, reviewing, and editing the manuscript.

\appendix

\section{Appendix}%{Synthetic Age--depth model}

\begin{figure}[htbp]	% Define function t
    \centering
    \includegraphics[width=\textwidth]{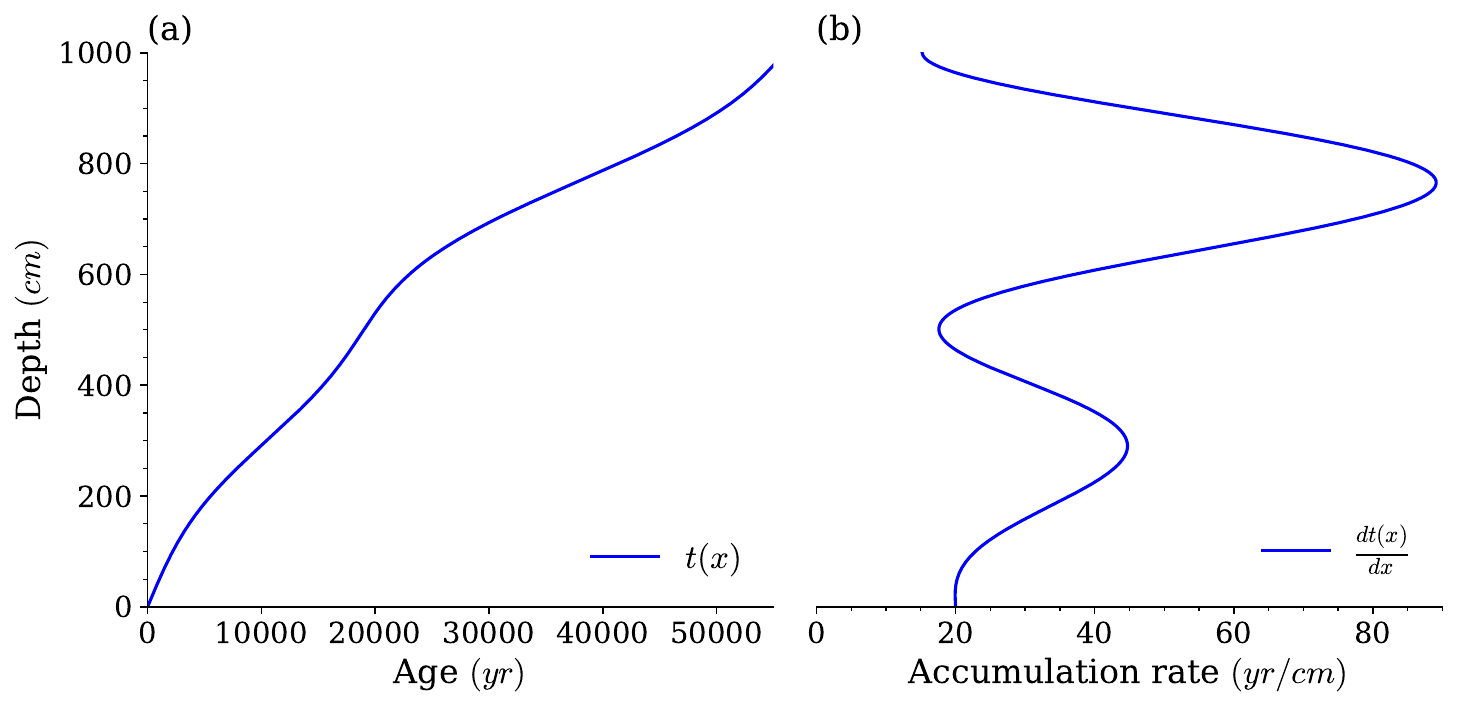}
    Synthetic age--depth function used to test \texttt{BSync}.  \caption{Panel (a) shows the behavior of a synthetic age--depth function $ t(x) $, calculated over a depth range of $x \in  [0, 1000]$ cm.  Panel (b) illustrates the corresponding derivative function, $dt(x)$, showing the change in deposition time with depth.
}
    \label{fig:combined_plots}
\end{figure}

Figure \ref{fig:combined_plots} presents the synthetic age--depth function $t(x)$ and its derivative 
$dt(x)$, which together illustrate the imposed linear trend and periodic variations in accumulation used to test the method.

\newpage
\section{Additional computational experiments}\label{appendix:additional_test}

To further expand on the comparison presented in Section~\ref{sec:bigmacs}, we conducted additional experiments to examine how \texttt{BSync} and \texttt{BIGMACS} respond to variations in proxy resolution and noise. These factors are often beyond the researcher's control but can have a significant impact on the reliability of alignment outputs.

\subsection{Impact of record resolution}

To evaluate model performance under varying sample sizes (i.e., record time resolution), we created downsampled versions of the synthetic record described in Section~\ref{sec:comp_exp}. We defined a set of sampling fractions, where 100\% corresponds to the full record. Each version retains a fixed fraction of the original observations, selected at uniform spacing across the series. For example, from a 100 observations, a 50\% record retains 50 observations at uniform sapcing.

\begin{figure}[htbp]
	\centering
	\includegraphics[width=0.95\textwidth]{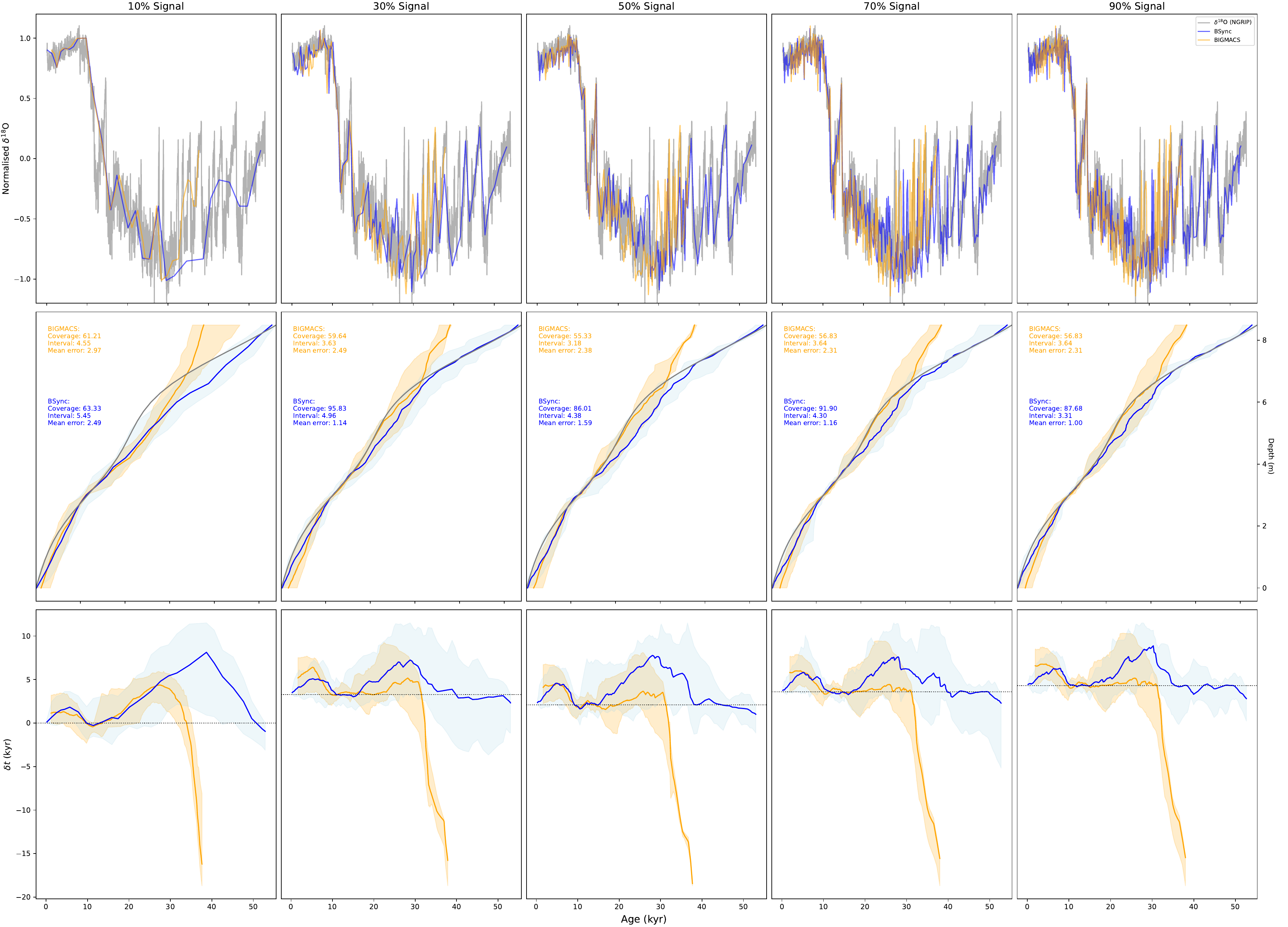}
  	\caption{
    Comparison of \texttt{BIGMACS} (orange) and \texttt{BSync} (blue) using different levels of signal percentage (left to right). The top panels show proxy synchronizations, the middle panels show the resulting age--depth relationship, and the bottom panels show the deviations from the true age expressed in units of standard deviations. The true age--depth relationship is shown in black.
    }
	\label{fig:6}
\end{figure}

Figure~\ref{fig:6} summarizes the effect of increasing sample size on coverage, accuracy, and interval length for both routines. In terms of coverage, \texttt{BSync} improves consistently with sample size, rising from 63.33\% in the most sparsely sampled case, to 91.90\% and 87.68\% in the two highest--resolution scenarios. By contrast, \texttt{BIGMACS} shows lower coverage throughout, ranging between 55.33\% and 61.21\%, with no evident improvement as sample size increases. For accuracy, \texttt{BSync} shows a marked reduction in mean error as sample size increases, decreasing from 2.49 kyr to 1.00 kyr. In contrast, \texttt{BIGMACS} shows only modest gains: mean error remains relatively flat around 2.3--2.9 kyr, improving marginally only in the highest--resolution cases. Finally, the two routines behave differently in interval length. \texttt{BSync} produces wider intervals in the lowest--resolution setting (5.45 kyr), which contracts steadily with increasing sample size to 3.31 kyr in the most densely sampled scenario. \texttt{BIGMACS}, on the other hand, yields interval lengths that are nearly constant across scenarios, ranging around 3.2--3.6 kyr, indicating limited sensitivity to increased sampling resolution.

%\begin{landscape}
%\end{landscape}

\subsection{Impact of noise and resolution}

We further evaluated how performance changes as the input--record noise increases. Higher noise can obscure proxy structure, making alignment more difficult and increasing the influence  of the priors relative to the data on the posterior distribution. To test this, we added Gaussian noise to the synthetic input signal at five levels: 0.01, 0.05, 0.1, 0.3, and 0.5, meaning that we added 1\%, 5\%, 10\%, 30\% and 50\% noise to the record . These experiments were repeated across multiple sampling densities to assess the joint effect of noise and resolution.

\begin{figure}[htbp]
	\centering \includegraphics[width=0.95\textwidth]{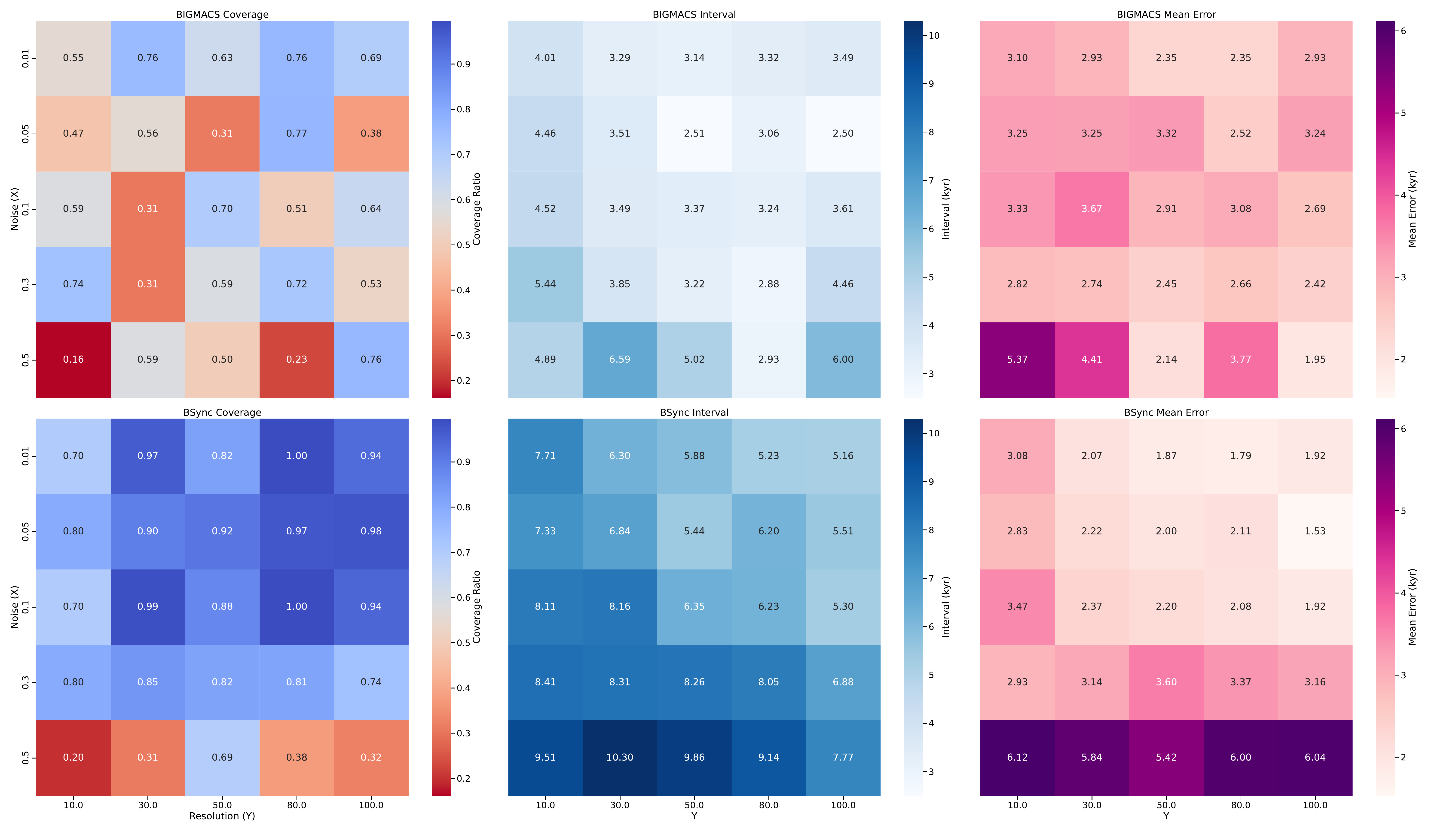}
  	\caption{Joint effects of proxy noise and sampling resolution on model performance. Each row compares \texttt{BIGMACS} (top) and \texttt{BSync} (bottom) across increasing noise levels (vertical axis) and coarser sampling resolution (horizontal axis). Evaluation metrics include coverage (left), mean interval length (center), and mean absolute error (right). \texttt{BSync} achieves consistently higher coverage and lower error across most conditions, using wider intervals adaptively when needed.}
	\label{fig:7}
\end{figure}

Figure~\ref{fig:7} presents the results as heatmaps, with the top row showing the results for \texttt{BIGMACS} and the bottom row for \texttt{BSync}. Each column correspond to adifferent evaluation metric: coverage, mean interval length, and mean absolute error. The results again suggest that \texttt{BIGMACS} struggles in many scenarios, particularly when records are both noisy and low--resolution, yielding low coverage and comparatively high error. By contrast, \texttt{BSync} is more robust, maintaining higher coverage and lower error across a broad range of settings. Although \texttt{BSync} widens its credible intervals under high--noise, low--resolution conditions, this adaptation preserves calibration and overall reliability.

%\begin{landscape}
%\end{landscape}

\end{document}